\documentclass[%
reprint,
superscriptaddress,
amsmath,amssymb,
]{revtex4-1}
\usepackage{graphicx}
\usepackage{dcolumn}
\usepackage{bm}
\usepackage{comment}
\usepackage[usenames,dvipsnames]{color}
\usepackage[mathlines]{lineno}
\DeclareGraphicsExtensions{.pdf,.png,.jpg}
\newcommand{\BBless}{\ensuremath{0\nu\beta\beta}}

\newcommand{\TeO}{\ensuremath{\rm TeO_{2}}}

\newcommand{\red}[1]{{\color{black}{#1}}}

\newcommand{\blue}[1]{{\color{black}{#1}}}
\newcommand{\ckky}{counts/(keV$\cdot$kg$\cdot$yr)}
\newcommand{\ky}{kg$\cdot$yr}
\begin{document}
\newcommand{\CombinedTeOneThirtyBBlessHalflife}{\red{$4.0\times 10^{24}~\mathrm{yr}$}}
\newcommand{\CombinedMbbRangeExclSMandPHFB}{\red{270\,--\,650~$\mathrm{meV}$}}  
\newcommand{\CombinedMbbRangeAllNME}{\red{270\,--\,760~$\mathrm{meV}$}}  
\newcommand{\CombinedTeOneThirtyBBlessHalflifeFreq}{\red{$4.1\times 10^{24}~\mathrm{yr}$}}
\newcommand{\QZeroTeOneThirtyBBlessHalflifeFreq}{\red{$2.8\times 10^{24}~\mathrm{yr}$}}
\newcommand{\TypicalPulseRiseTime}{\red{0.05~s}}
\newcommand{\TypicalPulseDecayTime}{\red{0.2~s}}
\newcommand{\PulseHeightmVPerkeV}{\red{$0.3~\mathrm{\mu V/keV}$}}
\newcommand{\TrigThresholdLower}{\red{30~keV}}
\newcommand{\TrigThresholdUpper}{\red{120~keV}}
\newcommand{\TriggerEfficencyFromHeaters}{\red{($98.529\pm 0.004$)\%}}
\newcommand{\PSAandPileUpEfficency}{\red{($93.7 \pm 0.7$)\%}}
\newcommand{\BBContainmentEfficiencyGFour}{\red{($88.35\pm 0.09$)\%}}
\newcommand{\AntiCoincidenceEfficiencyKForty}{\red{($99.64 \pm 0.10$)\%}}
\newcommand{\FinalTotalSelectionEfficiency}{\red{($81.3 \pm 0.6$)\%}}
\newcommand{\InstrumentedTeOTwoMass}{\red{38.3~kg}}
\newcommand{\FinalCUOREZeroExposure}{\red{35.2~kg$\cdot$yr}}
\newcommand{\FinalCUOREZeroIsoExposure}{\red{9.8~kg$\cdot$yr}}
\newcommand{\ExposureUsingDOFAndHeaterTGS}{\red{12\%}}
\newcommand{\ExposureUsingDOFAndCalibrationTGS}{\red{8\%}}
\newcommand{\ExposureUsingStandardOFAndCalibrationTGS}{\red{21\%}}
\newcommand{\MixedEstimatorResolutionGain}{\red{4\%}}
\newcommand{\MixedEstimatorExposureGain}{\red{12\%}}
\newcommand{\ExposureLossFromBadIntervals}{\red{7\%}}
\newcommand{\QBBTeOneThirty}{\red{$2527.518 \pm 0.013$~keV}}
\newcommand{\NatAbundanceTeOneThirty}{\red{$34.167\%$}}
\newcommand{\QinoExposureTeOneThirty}{\red{19.75~{\ky}}}
\newcommand{\QinoROIBkgIndex}{\red{$ 0.169 \pm 0.006$~\ckky}}
\newcommand{\QinoMeanEnergyResolution}{\red{6.9~keV}}
\newcommand{\QinoRMSEnergyResolution}{\red{2.9~keV}}
\newcommand{\CobaltPeakBestFitPosition}{\red{$2507.6 \pm 0.7$~keV}}
\newcommand{\CobaltEnergyOffset}{\red{$1.9\pm 0.7$~keV}}
\newcommand{\SingleEscapeEnergyOffset}{\red{$0.84\pm 0.22$~keV}}
\newcommand{\FinalCalibrationResidual}{\red{0.12~keV}}
\newcommand{\BackgroundResolutionScalingFactor}{\red{$1.05 \pm 0.05$}}
\newcommand{\BackgroundResolutionCorrectionInFit}{\red{1.05}}
\newcommand{\BackgroundResolutionCorrectionSystematic}{\red{\pm0.05}}
\newcommand{\NumberOfEventsInROI}{\red{233}}
\newcommand{\QZeroOnlyBBlessDecayRate}{\red{$0.01 \pm 0.12\,(\mathrm{stat.})\pm 0.01\,(\mathrm{syst.})\times 10^{-24}~\mathrm{yr}^{-1}$}}
\newcommand{\QZeroFinalBackgroundIndex}{\red{$0.058 \pm 0.004\,(\mathrm{stat.})\pm 0.002\,(\mathrm{syst.})$~\ckky}} 
\newcommand{\QZeroBBlessRateULNinetyStatOnly}{\red{$\Gamma_{0\nu} < 0.25 \times 10^{-24}~\mathrm{yr}^{-1}$}}. 
\newcommand{\QZeroBBlessRateULNinetyWithSyst}{\red{$\Gamma_{0\nu}< 0.25 \times 10^{-24}~\mathrm{yr}^{-1}$}}. 
\newcommand{\QZeroOnlyBBlessHalflifeLLWithSyst}{\red{$ 2.7\times 10^{24}~{\rm yr}$}}
\newcommand{\QZeroOnlyBBlessHalflifeLLStatonly}{\red{$ 2.7\times 10^{24}~{\rm yr}$}}
\newcommand{\QZeroBBlessLLSensitivity}{\red{$2.9\times 10^{24}~{\rm yr}$}}
\newcommand{\QZeroBetterLimitProbability}{\red{54.7\%}}
\newcommand{\NumberOfEventsInToyMC}{\red{233}}
\newcommand{\BBlessShiftFromEnergyScale}{\red{0.006}}
\newcommand{\ScalingBBlessShiftFromEnergyScale}{\red{0.4}}
\newcommand{\BBlessShiftFromEnergyResolution}{\red{0.006}}
\newcommand{\ScalingBBlessShiftFromEnergyResolution}{\red{2.6}}
\newcommand{\BBlessShiftFromLineShape}{\red{0.004}}
\newcommand{\ScalingBBlessShiftFromLineShape}{\red{1.3}}
\newcommand{\BBlessShiftFromBkgShape}{\red{0.004}}
\newcommand{\ScalingBBlessShiftFromBkgShape}{\red{0.7}}
\newcommand{\BBlessShiftFromFitBias}{\red{0.006}}
\newcommand{\BBlessShiftFromFitBiasScaling}{\red{0.15}}
\newcommand{\BBlessSystErrorFromCutEfficiencies}{\red{0.7\%}}
\newcommand{\GoodnessOfFitChiSq}{\red{90\%}}
\newcommand{\QZeroEffectiveResolutionRMSWeighted}{\red{$2.9~{\rm keV}$}} 
\newcommand{\EffectiveFWHMAtQbbBackgroundData}{\red{$5.1\pm 0.3{\rm~keV}$}}

\title{Search for Neutrinoless Double-Beta Decay of $^{130}$Te with CUORE-0}
\newcommand{\deceased}{\altaffiliation {Deceased.}}
\newcommand{\atyalenow}{\altaffiliation {Present address: Department of Physics, Yale University, New Haven, Connecticut 06520, USA}}
\newcommand{\atprincetonnow}{\altaffiliation {Present address: Physics Department, Princeton University, Princeton, NJ 08544, USA}}

\author{K.~Alfonso}
\affiliation{ Department of Physics and Astronomy, University of California, Los Angeles, CA 90095 - USA }

\author{D.~R.~Artusa}
\affiliation{ Department of Physics and Astronomy, University of South Carolina, Columbia, SC 29208 - USA }
\affiliation{ INFN - Laboratori Nazionali del Gran Sasso, Assergi (L'Aquila) I-67010 - Italy }

\author{F.~T.~Avignone~III}
\affiliation{ Department of Physics and Astronomy, University of South Carolina, Columbia, SC 29208 - USA }

\author{O.~Azzolini}
\affiliation{ INFN - Laboratori Nazionali di Legnaro, Legnaro (Padova) I-35020 - Italy }

\author{M.~Balata}
\affiliation{ INFN - Laboratori Nazionali del Gran Sasso, Assergi (L'Aquila) I-67010 - Italy }

\author{T.~I.~Banks}
\affiliation{ Department of Physics, University of California, Berkeley, CA 94720 - USA }
\affiliation{ Nuclear Science Division, Lawrence Berkeley National Laboratory, Berkeley, CA 94720 - USA }

\author{G.~Bari}
\affiliation{ INFN - Sezione di Bologna, Bologna I-40127 - Italy }

\author{J.W.~Beeman}
\affiliation{ Materials Science Division, Lawrence Berkeley National Laboratory, Berkeley, CA 94720 - USA }

\author{F.~Bellini}
\affiliation{ Dipartimento di Fisica, Sapienza Universit\`{a} di Roma, Roma I-00185 - Italy }
\affiliation{ INFN - Sezione di Roma, Roma I-00185 - Italy }

\author{A.~Bersani}
\affiliation{ INFN - Sezione di Genova, Genova I-16146 - Italy }

% ...

\collaboration{CUORE Collaboration}
\date{\today}
\begin{abstract}
We report the results of a search for neutrinoless double-beta decay
in a {\FinalCUOREZeroIsoExposure} exposure of $^{130}$Te using a
bolometric detector array, CUORE-0.  
The characteristic detector energy resolution and background level in the region of interest 
are {\EffectiveFWHMAtQbbBackgroundData}\,FWHM and
{\QZeroFinalBackgroundIndex}, respectively.  The median 90\,\%~C.L.\ lower-limit half-life sensitivity of the experiment is
{\QZeroBBlessLLSensitivity} and surpasses the sensitivity of previous searches.
We find no evidence for neutrinoless
double-beta decay of $^{130}$Te and place a Bayesian lower bound on the decay
half-life, $T^{0\nu}_{1/2}>$~{\QZeroOnlyBBlessHalflifeLLWithSyst} at 90\,\%~C.L.
Combining CUORE-0 data with the {\QinoExposureTeOneThirty}~exposure of $^{130}$Te from the Cuoricino 
experiment we obtain $T^{0\nu}_{1/2} > $~{\CombinedTeOneThirtyBBlessHalflife} at
90\,\%~C.L.~(Bayesian), the most stringent limit to date on this half-life. 
Using a range of nuclear matrix element estimates we interpret this as a limit on the effective Majorana neutrino mass,
$m_{\beta\beta}<$~{\CombinedMbbRangeAllNME}.
\end{abstract}
\pacs{Valid PACS appear here}
\maketitle
Neutrinoless double-beta (\BBless)~decay is a hypothesized
lepton-number-violating process~\cite{Pontecorvo:1967fh} that has
never been decisively observed. Its discovery would 
prove that lepton number is not a symmetry of nature, establish that neutrinos
are Majorana fermions, possibly constrain the absolute neutrino mass
scale, and support theories that leptons seeded the  matter-antimatter asymmetry in the universe~\cite{Luty:1992un}. 
The clear potential for fundamental impact has motivated intense 
effort to search for this
decay~\cite{bib:AGiuliani,bib:OCremonesi,Barabash:2014bfa}.

The Cryogenic Underground Observatory for Rare Events
(CUORE)~\cite{Arnaboldi:2002du, Ardito:2005ar}, now in the final
stages of construction at Laboratori Nazionali del Gran
Sasso~(LNGS), promises to be one of the most sensitive
upcoming \BBless~decay searches.  The detector exploits the bolometric
technique~\cite{Fiorini:1983yj,bib:bolorev} in $\rm 5\times 5\times 5~cm^{3}$
$^{\rm nat}${\TeO} crystals, whereby the tiny heat capacity attained by a crystal at $\sim$10~mK results in a measurable increase of its temperature when it absorbs energy. 
The sought-after signature of {\BBless}~decay
is a peak in the measured energy spectrum at the transition energy ($Q_{\beta\beta}$), which
for $^{130}$Te is {\QBBTeOneThirty}~\cite{Redshaw:2009zz}. 

CUORE will consist of 19 towers containing 52 crystals each;
\mbox{CUORE-0} is 
one such tower built using
the low-background assembly techniques developed for
CUORE~\cite{Buccheri:2014bma}.
The 52
crystals~\cite{Arnaboldi:2010fj} are held
in an ultra-pure copper frame by polytetrafluoroethylene
supports and arranged in 13 floors, with 4 crystals per floor. 
Each crystal is instrumented with a neutron-transmutation-doped Ge
thermistor~\cite{bib:HallerNTD} to record thermal pulses and a silicon
heater to generate reference
pulses~\cite{Andreotti:2012zz}.
The tower is deployed in Hall~A of LNGS and exploits the cryogenic
system, shielding configuration, and electronics
from a predecessor experiment,
Cuoricino~\cite{Arnaboldi:2004qy,Arnaboldi:2008ds,Andreotti:2010vj}.

\mbox{CUORE-0} represents the state of the art for large-mass, low-background, ultra-low-temperature bolometer
arrays.  
While also a competitive {\BBless}~decay search, it has validated the ultraclean assembly 
techniques and radiopurity of materials for the upcoming CUORE experiment. Technical details can be found in~\cite{Arnaboldi:2010fj,Alessandria:2011vj,Alessandria:2012zp,Buccheri:2014bma, bib:detectorPaper}; we focus here on the first physics results from CUORE-0.

The data were collected in twenty month-long blocks called
\emph{datasets} during two campaigns which ran
from March 2013 to August 2013 and from November 2013 to March 2015.
For approximately three days at the beginning and
end of each dataset we calibrated the detector by placing thoriated
wires next to the outer vessel of the cryostat.  
Data collected between calibrations, denoted \emph{physics data}, are used for the {\BBless}~decay search.

Each thermistor voltage, except for one thermistor which we failed to wire bond,  is continuously acquired at a rate of $125$~Hz.  
Events are identified using a software trigger with a channel-dependent threshold of between {\TrigThresholdLower}~and {\TrigThresholdUpper}. 
\blue{The trigger rate per bolometer is 60~mHz (1~mHz) in calibration (physics) mode.}  
Particle-induced pulses have rise (decay) times of $\sim${\TypicalPulseRiseTime} ($\sim${\TypicalPulseDecayTime}), and 
have amplitudes of $\sim${\PulseHeightmVPerkeV} before amplification.  
We analyze a 5-s-long window consisting of 1~s before and 4~s after each trigger.  The pre-trigger voltage establishes the bolometer temperature before the event; the pulse amplitude
establishes the event energy.  Every 300~s, a 
stable current pulse is injected in each heater to generate tagged monoenergetic reference pulses. 
Noise waveforms are collected on all bolometers every 200~s.

The analysis utilizes two 
pulse-filtering techniques, denoted optimal filter~(OF) and decorrelated optimal
filter~(DOF), 
and two methods for thermal gain stabilization~(TGS), denoted heater-TGS and calibration-TGS. 
The filters optimize energy resolution \cite{bib:OF} by exploiting the distinct frequency characteristics of particle-induced vs.\ noise pulses.  TGS corrects for small changes in the energy-to-amplitude response of the detection chain
using monoenergetic heater or calibration events.
Both OF and heater-TGS were used for Cuoricino~\cite{Andreotti:2010vj}.
We developed DOF to reduce correlated noise between adjacent crystals; such noise mainly
affects the upper floors of the tower closest to cryostat noise sources~\cite{ManciniTerracciano:2012fq,bib:JonsThesis}. 

To recover data from the two bolometers with non-functioning heaters and from periods when temperature drifts in a bolometer exceeded the linear dynamic range of the heater-TGS, we developed 
calibration-TGS, which uses the 2615~keV $^{208}$Tl calibration line.  To successfully apply calibration-TGS to physics data, we 
monitor 
parameters that can affect the 
bolometer response between calibrations (e.g., drifts in DC offset or amplifier gain).   Where possible we employ both TGS methods, 
yielding up to four stabilized pulse-amplitude estimators for each event~(OF and DOF, with heater- and calibration-TGS). 

\begin{figure}[]
\centering
\includegraphics[width=0.5\textwidth]{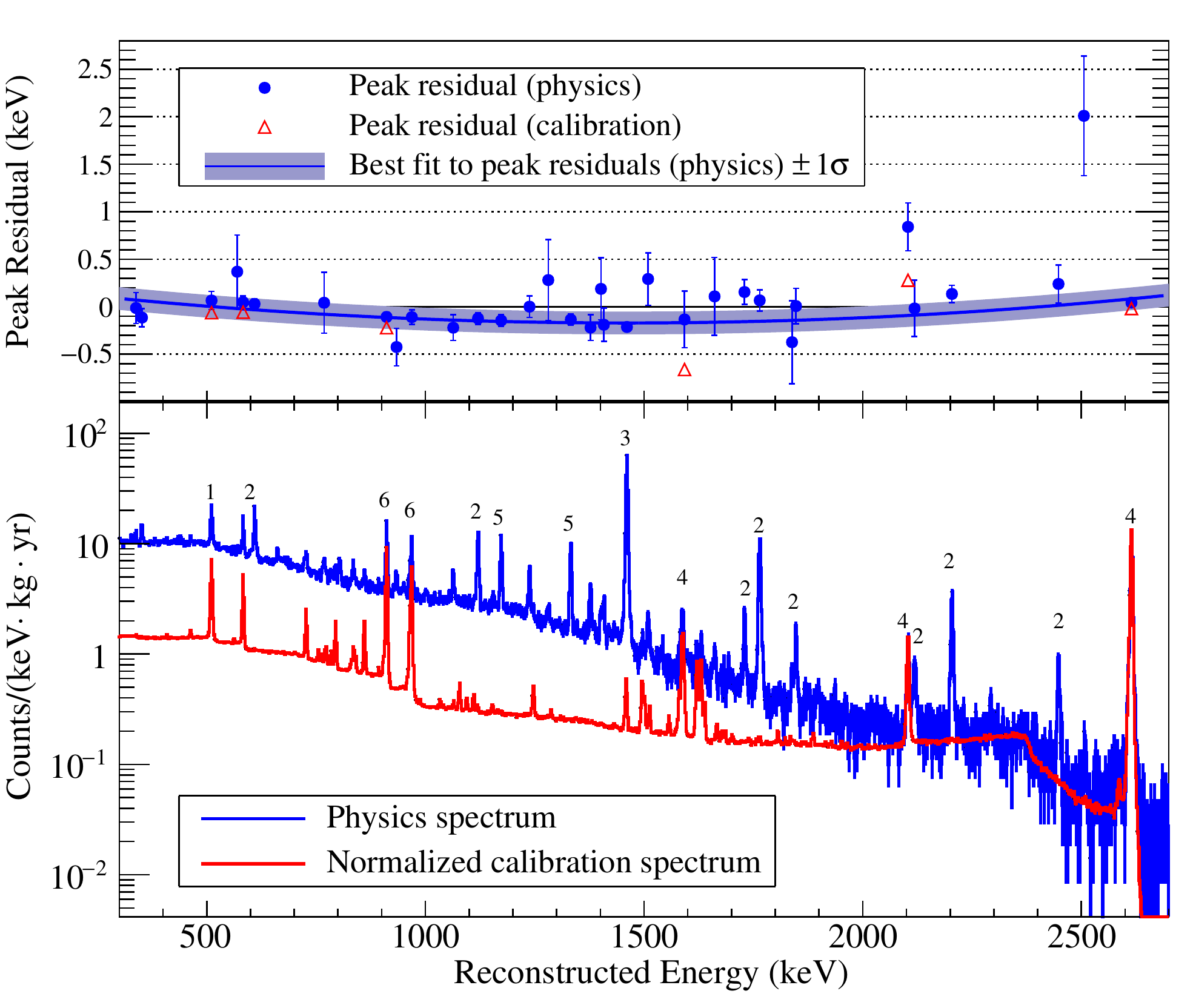}
\caption{ Bottom: Energy spectra of physics (blue) and calibration (red) data; the latter is normalized relative to the former at 2615~keV. The peaks are identified as: (1)~$e^{+}e^{-}$ annihilation, (2)~$^{214}$Bi, (3)~$^{40}$K, (4)~$^{208}$Tl, (5)~$^{60}$Co, and (6)~$^{228}$Ac.  Top: 
Difference of best-fit reconstructed peak energy and expected peak-energy for physics (blue points) and calibration (red) data. The blue line is the best-fit function to the physics peak residuals; the shaded band is its 1$\sigma$ uncertainty.}
\label{fig:gamma_lines}
\end{figure}

To convert these to energy, we correlate prominent peaks in the stabilized-amplitude spectra collected in calibration runs with gamma lines of known energy between 511~keV and 2615~keV (Fig.~\ref{fig:gamma_lines}).  We fit a quadratic function with zero intercept to the peak-mean vs.\ known-energy points to determine a calibration function for each stabilized-amplitude estimator of each bolometer-dataset and apply these to the physics data.  
To avoid biasing the subsequent analysis we then {\emph{blind}} the physics data in the region of interest~(ROI) using a procedure \cite{Aguirre:2014lua} which produces an 
artificial peak at $Q_{\beta\beta}$.  


We select the best-performing energy estimator for each bolometer-dataset to optimize sensitivity to \BBless~decay (quantified by the ratio of energy resolution of the 2615~keV calibration line to the physics data exposure).  While the combination of OF with heater-TGS is the default choice, 
combinations involving DOF and calibration-TGS --- which are more robust against low-frequency common-mode noise and long-term temperature drifts, respectively --- are selected 
if the improvement relative to the default is statistically significant.  
The fractions of exposure using OF with calibration-TGS, DOF with heater-TGS, and DOF with calibration-TGS are~{\ExposureUsingStandardOFAndCalibrationTGS}, {\ExposureUsingDOFAndHeaterTGS}, and {\ExposureUsingDOFAndCalibrationTGS}, respectively.  These new techniques result in a {\MixedEstimatorResolutionGain}~improvement in energy resolution and a {\MixedEstimatorExposureGain}~increase in usable exposure. 

We select {\BBless}~decay candidates 
in the physics data
according to the following conditions.
First, we discard 
low-quality data (e.g., periods of cryostat instability or equipment malfunction), reducing the total exposure by {\ExposureLossFromBadIntervals}.
To allow a bolometer time to equilibrate after each event (pileup rejection) we require that the times since the previous event and until the next event on the same bolometer are greater than 3.1~s and 4.0~s, respectively. To reject noisy pulses which can contribute to background we require each waveform to be consistent with a reference waveform, constructed for each bolometer-dataset from calibration data around the 2615~keV $^{208}$Tl peak.  Six pulse-shape parameters characterize the waveforms, and the acceptance criteria are tuned
simultaneously on prominent peaks in the physics data to maximize the signal sensitivity at each peak. These peaks range in energy between 146~keV and 2615~keV.  The sensitivity is quantified by the ratio of signal accepted to square-root of background accepted, where the signal sample is drawn from events that populate each peak and the background is drawn from nearby off-peak events. 
The tuning uses 50\% of the data, randomly selected, and excludes the ROI.  
To reduce background from 
decays depositing energy in multiple crystals (e.g., $\alpha$'s
 at  crystal surfaces or multiple Compton scatters) we reject an event if another occurs in the tower within $\pm$5~ms (anticoincidence).  

The selection efficiencies are evaluated with the fraction of data not used for 
tuning and averaged over all bolometer-datasets.  The trigger efficiency is
estimated from the fraction of heater pulses that produce an event trigger; we also exploit the heater events to measure the energy reconstruction efficiency (i.e., the probability for a monoenergetic pulse to reconstruct correctly). The combined trigger and reconstruction efficiency is {\TriggerEfficencyFromHeaters}.
The combined efficiency of the pileup and pulse-shape selection, estimated from the fraction of 2615~keV $^{208}$Tl events in physics data that pass this selection,  is {\PSAandPileUpEfficency}.  
The anticoincidence efficiency has two components: the probability for a {\BBless}~decay to be fully
contained in one 
crystal and the probability for it to survive accidental coincidences. The former, estimated from simulation~\cite{Agostinelli:2002hh}, is {\BBContainmentEfficiencyGFour}; the latter we 
find to be {\AntiCoincidenceEfficiencyKForty} using the
1461~keV $\gamma$-ray 
from $^{40}$K\,.
%
The total selection efficiency is {\FinalTotalSelectionEfficiency}.

We use the high-statistics 2615~keV $^{208}$Tl line in calibration data to establish the detector response to a monoenergetic deposit (lineshape) near the ROI. The data exhibit a slightly non-Gaussian lineshape characterized by a primary peak and a secondary peak whose mean is lower in energy by $\sim$0.3\% and whose amplitude is typically $\sim$5\% of the primary peak. \blue{Non-Gaussian low-energy structure was also observed in Cuoricino~\cite{bib:AdamsThesis,bib:CarrettonisThesis}}.  The origin of this structure in CUORE-0 is 
under investigation.  We studied several lineshapes, including double- and triple-Gaussian models; while the latter perform well at the $^{208}$Tl line, we adopt the double-Gaussian lineshape as it is the simplest 
that reproduces the detector response over the broadest energy range.  
\begin{figure}[t]
\centering 
\includegraphics[width=0.5\textwidth]{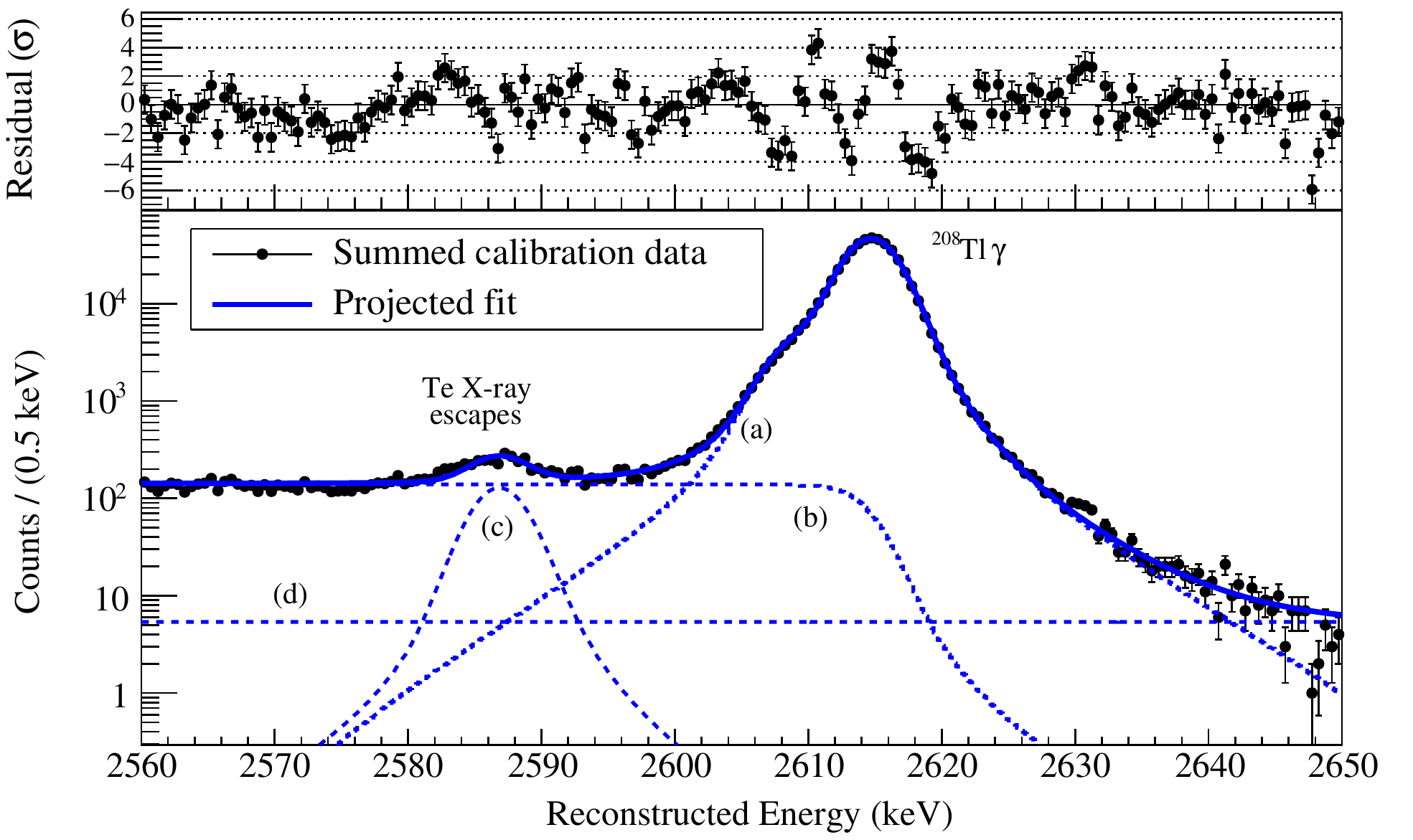} 
\caption{\blue{Bottom: Calibration data near the
  2615~keV $^{208}$Tl $\gamma$-ray line, 
integrated over 
all bolometer-datasets. The solid blue line is the projection of the UEML fit described in the main text.  In addition to the double-Gaussian lineshape for each bolometer-dataset, the fit function 
includes terms to model a multiscatter Compton continuum, a $\sim$\,30~keV Te X-ray escape peak, and a continuum background; these components, summed over all bolometer-datasets, are indicated by the blue dashed lines (a), (b), (c), and (d), respectively.
Top: Normalized residuals of the data and the best-fit model.}}
\label{fig:resolution}
\end{figure}

We parametrize the lineshape $\rho$ for each bolometer-dataset ($b$,~$d$) as $\rho_{b,d} = \rho(\mu_{b,d},\sigma_{b,d}, \delta_{b,d}, \eta_{b,d})$. For each ($b$,~$d$) pair, $\mu_{b,d}$ is the mean of the primary peak, $\delta_{b,d}$ is the ratio of the means of the secondary and primary peaks, $\sigma_{b,d}$ is the common Gaussian width of both peaks, and $\eta_{b,d}$ is the fractional intensity of the secondary peak.  We estimate these parameters
with a simultaneous, unbinned extended maximum likelihood~(UEML) fit to the 2615~keV $^{208}$Tl calibration line 
(Fig.~\ref{fig:resolution}); the resulting best-fit parameters are denoted $\hat{\mu}_{b,d}$, $\hat{\sigma}_{b,d}$, $\hat{\delta}_{b,d}$, and $\hat{\eta}_{b,d}$.

We next repeat 
this lineshape fit on a series
of peaks of
known energy between 511~keV and 2615~keV in the physics data (Fig.~\ref{fig:gamma_lines}). 
For a peak of known energy $E$,  $\mu_{b,d}(E)$ can vary
around the expected calibrated energy via a single free parameter $\Delta \mu (E)$. 
To 
treat energy dependence of the resolution or possible differences in resolution between calibration vs.\ physics data,  we vary the $\sigma_{b,d}$ relative to $\hat{\sigma}_{b,d}$ via a global scaling parameter $\alpha_{\sigma}(E)$.  \blue{We fix the $\delta_{b,d}$ and $\eta_{b,d}$ to the corresponding $\hat{\delta}_{b,d}$ and $\hat{\eta}_{b,d}$.}  

The energy residual parameters $\Delta \mu(E)$ are plotted in Fig.~\ref{fig:gamma_lines}.
A prominent outlier is the peak attributed to $^{60}$Co double-gamma events which reconstructs at {\CobaltPeakBestFitPosition}, {\CobaltEnergyOffset} higher than 
expected~\cite{bib:NNDC}; \blue{a shift of $0.8 \pm 0.3$~keV was observed in Cuoricino~\cite{bib:AdamsThesis}}.   The single-escape peak of the $^{208}$Tl 2615~keV gamma at 2104~keV also reconstructs higher by {\SingleEscapeEnergyOffset}.  
Data taken with a $^{60}$Co source confirm the double-gamma events reconstruct at higher energy, in agreement with our physics data.   Simulations show their
energy deposit in a bolometer is 
less localized than the single-gamma lines 
studied; this may be responsible for the observed response.
The double-escape peak of the $^{208}$Tl 2615~keV line ($E \simeq 1593$~keV) reconstructs within $0.13\pm 0.30$~keV of the expected value. Since $e^{+}e^{-}$ pairs and {\BBless}~decays share similar event topologies we assume the latter 
would reconstruct according to the calibrated energy scale.
%

We estimate the calibration offset at $Q_{\beta\beta}$ from a parabolic fit to the physics-peak residuals in Fig.~\ref{fig:gamma_lines}, excluding the $^{60}$Co double-gamma and $^{208}$Tl single-escape lines as outliers. We adopt the standard deviation of the parabolic-fit residuals as a systematic uncertainty. The result is \blue{\mbox{$\Delta \mu(Q_{\beta\beta}) = 0.05 \pm 0.05 \mathrm{(stat.)} \pm 0.09 \mathrm{(syst.)}$~keV.}}  Similarly, fitting the resolution-scaling parameters with a linear function we find $\alpha_{\sigma}(Q_{\beta\beta}) = 1.05 \pm 0.05$. 
\blue{Using this $\alpha_{\sigma}(Q_{\beta\beta})$, we estimate from calibration data the FWHM at $Q_{\beta\beta}$ of each bolometer-dataset in physics data.  We quote the exposure-weighted harmonic mean of these physics FWHM values,  {\EffectiveFWHMAtQbbBackgroundData}, as a characteristic value of the detector resolution in the ROI~\cite{bib:JonsThesis}.  The RMS of the calibration FWHM values is {\QZeroEffectiveResolutionRMSWeighted}.}
\begin{figure}[t]
\centering 
\includegraphics[width=0.5\textwidth]{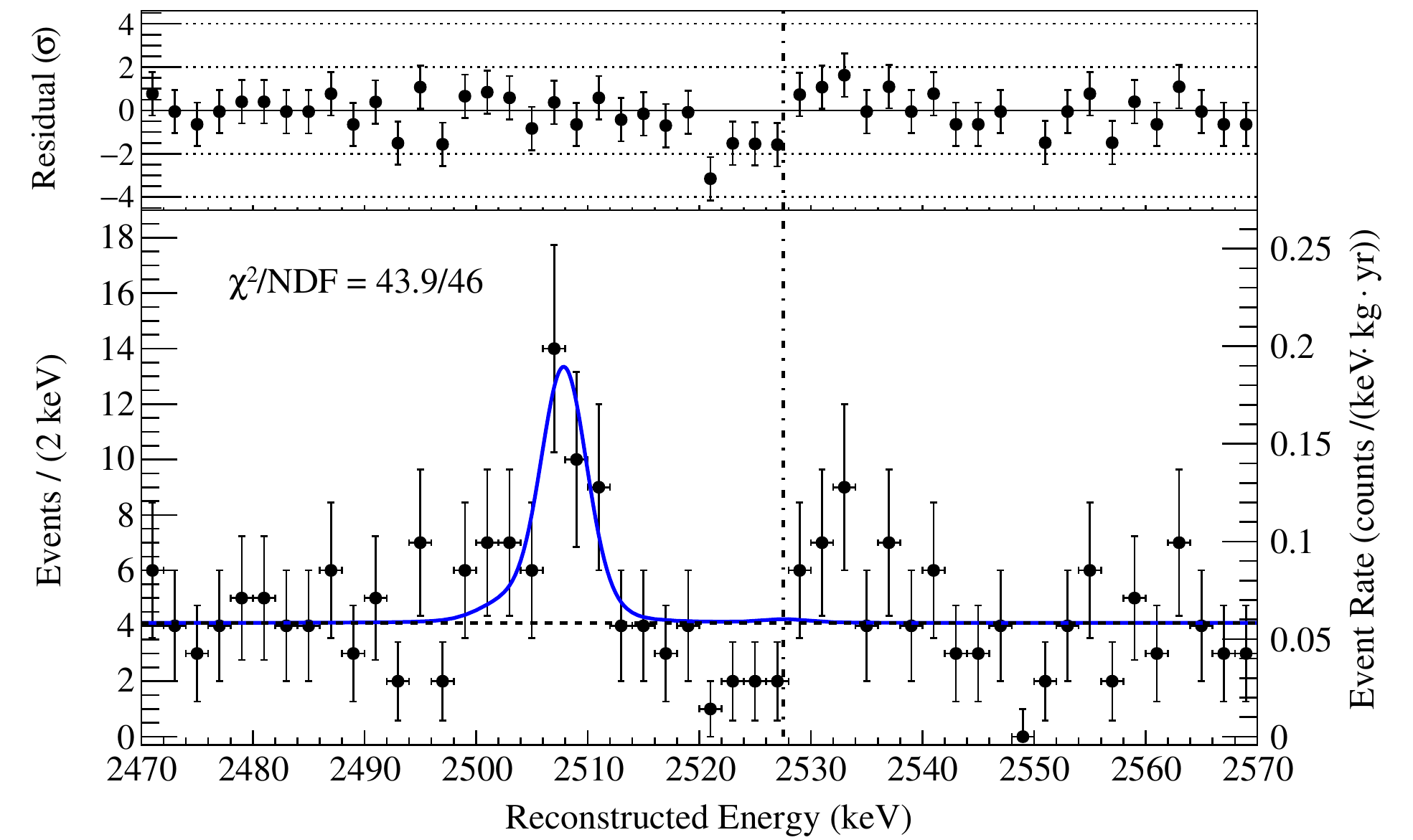} 
\caption{ \blue{Bottom: The best-fit model from the UEML fit 
(solid blue line) overlaid on the 
spectrum of \BBless~decay candidates in CUORE-0 (data points); 
the data are shown with Gaussian error bars.  The peak at $\sim$2507~keV is attributed to $^{60}$Co; the dotted black line shows the continuum background component of the best-fit model.  Top: The normalized residuals of the best-fit model and the binned data. The vertical dot-dashed black line indicates the position of 
$Q_{\beta\beta}$.}} 
\label{fig:UnblindedSpectrum}
\end{figure}

After unblinding the ROI by removing the artificial peak, we determine the yield of {\BBless} decay events from a simultaneous UEML
fit~\cite{bib:AdamsThesis} in the energy region $2470$--$2570$~keV
(Fig.~\ref{fig:UnblindedSpectrum}).  
The fit components are: a posited signal peak at $Q_{\beta\beta}$, a peak at $\sim$\,2507~keV from $^{60}$Co double-gammas, and a 
 continuum background attributed to multiscatter Compton events from $^{208}$Tl and surface decays~\cite{bib:backgroundPaper}.  We model both peaks using the established lineshape.  For {\BBless}~decay, the  $\mu_{b,d}(Q_{\beta\beta})$ are fixed at the expected position (i.e.,  $87.00~\mathrm{keV} - \Delta \mu(Q_{\beta\beta})$ below $\hat{\mu}_{b,d}$, where $87.00$~keV is the nominal energy difference between $Q_{\beta\beta}$ and the $^{208}$Tl line), the $\sigma_{b,d}$ are fixed to be $1.05\times\hat{\sigma}_{b,d}$, the $\delta_{b,d}$ and $\eta_{d,b}$ are fixed to their best-fit calibration values, and the {\BBless}~decay rate ($\Gamma_{0\nu}$) is treated as a global free parameter.  The $^{60}$Co peak is treated in a similar way except that a global free parameter is added to the expected $\mu_{b,d}$ to accomodate the anomalous double-gamma reconstruction. The $^{60}$Co yield, although a free parameter, is constrained to follow the isotope's half-life~\cite{bib:NNDC} since it was cosmogenically produced above ground but is not replenished under ground at LNGS.  Within the limited statistics the continuum background can be modeled with a zeroth-order polynomial; we consider first- and second-order alternatives later.  

The ROI contains {\NumberOfEventsInROI} candidates
in a total exposure of {\FinalCUOREZeroExposure} of TeO$_2$, or {\FinalCUOREZeroIsoExposure} of $^{130}$Te considering the natural isotopic abundance of {\NatAbundanceTeOneThirty}~\cite{bib:TeAbundance}. 
The best-fit $\Gamma_{0\nu}$ is \mbox{{\QZeroOnlyBBlessDecayRate}}, and the best-fit background index in the ROI is \mbox{{\QZeroFinalBackgroundIndex}}. 
\begin{figure}[tp]
\centering 
\includegraphics[width=0.5\textwidth]{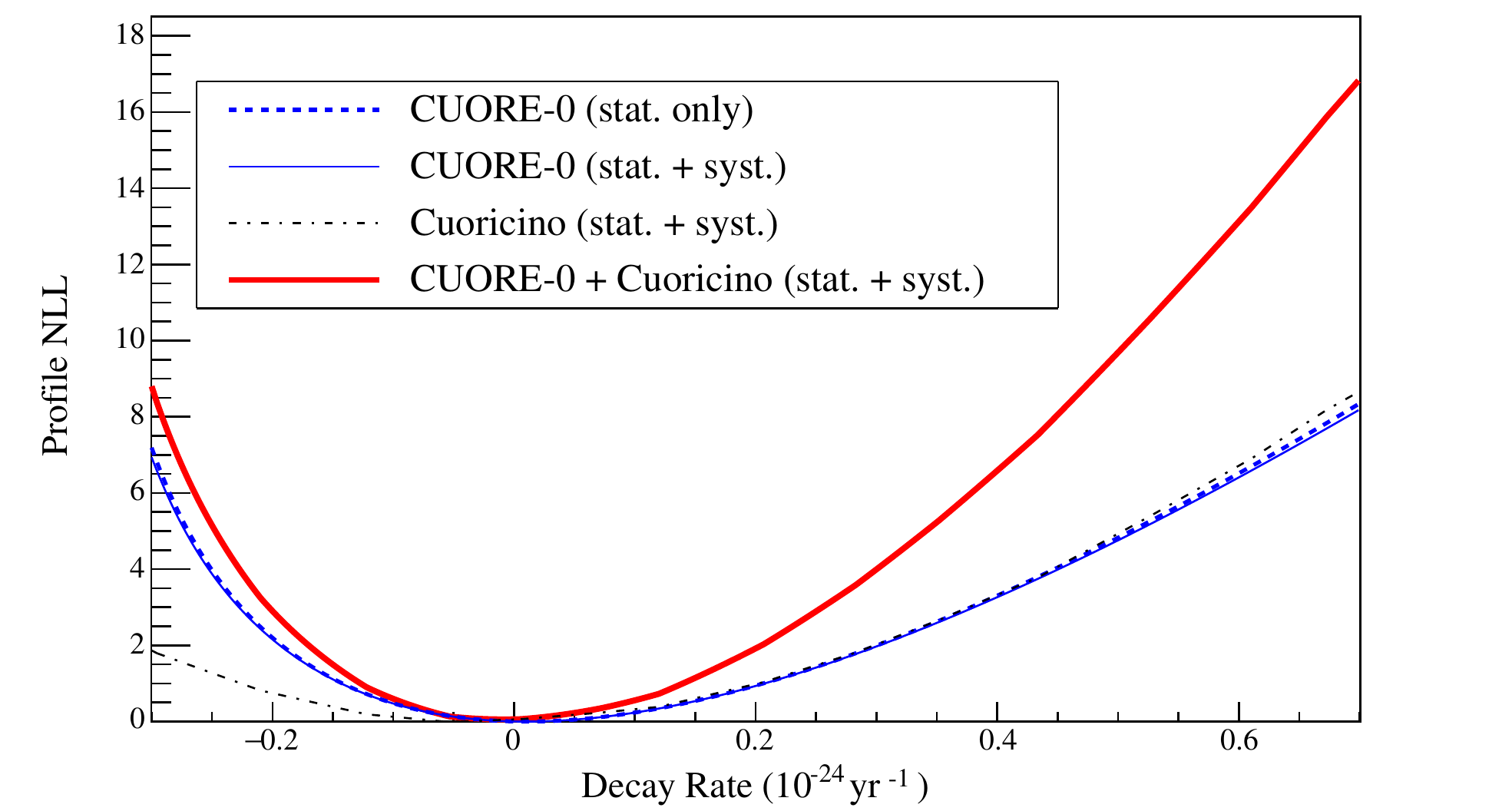} 
\caption{Profile negative log-likelihood (NLL) curves for CUORE-0, Cuoricino~\cite{Arnaboldi:2004qy,Arnaboldi:2008ds,Andreotti:2010vj}, and their combination.}
\label{fig:Profile_NLL}
\end{figure}

We evaluate the goodness of fit by comparing the value of the binned $\chi^2$ in Fig. \ref{fig:UnblindedSpectrum} (43.9 for 46 degrees of freedom) with the distribution from a large set of pseudo-experiments with 233 Poisson-distributed events in each, and generated with the best-fit values of all parameters; {\GoodnessOfFitChiSq} of 
trials return $\chi^2>43.9$.  The data are also compatible with this set of pseudo-experiments according to the Kolmogorov-Smirnov metric.  We quantify the significance of each of the positive and negative fluctuations about the best-fit function by comparing the likelihood of our best-fit model to the likelihood from an UEML fit where the fluctuation is modeled with a signal peak.  For one degree of freedom, the most negative (positive) fluctuation has a probability of 0.5\% (3\%). 
The probability to realize the largest observed fluctuation anywhere in the 100-keV ROI is $\sim10\%$.

%
We find no evidence for $0\nu\beta\beta$ decay 
and set a 90\% C.L. Bayesian upper limit at
{\QZeroBBlessRateULNinetyStatOnly}, or ~$T_{1/2}^{0\nu}>~${\QZeroOnlyBBlessHalflifeLLStatonly} (statistical uncertainties only); the prior used was uniform ($\pi(\Gamma_{0\nu})=1$ for $\Gamma_{0\nu}>= 0$). The median 90\% C.L.\ lower-limit sensitivity for $T_{1/2}^{0\nu}$ is {\QZeroBBlessLLSensitivity}. The probability to obtain a more stringent limit than the one reported above is
{\QZeroBetterLimitProbability}.  Including systematic uncertainties (Table~\ref{tab:sysError}) the 90\% C.L.\ limits are {\QZeroBBlessRateULNinetyWithSyst} or $T_{1/2}^{0\nu}>$~{\QZeroOnlyBBlessHalflifeLLWithSyst}.

\begin{figure}[tp]
\centering 
\includegraphics[width=0.5\textwidth]{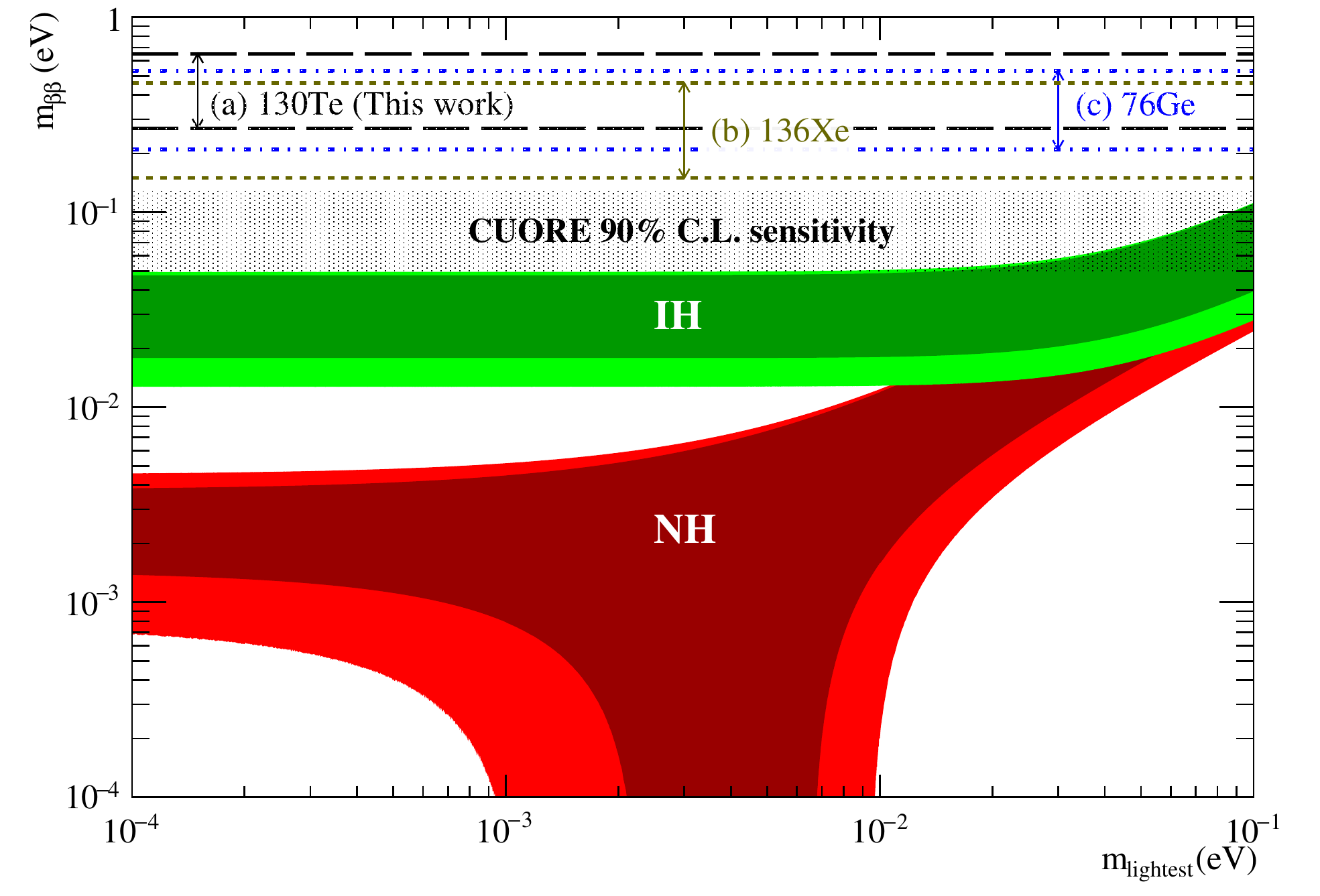} 
\caption{Constraints on $m_{\beta\beta}$ vs.\ lightest neutrino mass ($m_{\mathrm{lightest}}$). For the inverted (IH, green) and normal (NH, red) hierarchies the central dark band is derived from the best-fit neutrino oscillation parameters, the lighter outer band includes their 3$\sigma$ uncertainties~\cite{pdg:2014}.  The horizontal bands delineated by the long-dashed black lines (a), the dashed beige lines (b), and the dot-dashed blue lines (c) are the range of 90\% C.L upper limits on $m_{\beta\beta}$ coming from (a)$^{130}$Te (CUORE-0 combined with Cuoricino), (b)$^{136}$Xe (EXO-200~\cite{Albert:2014awa}, KamLAND-Zen~\cite{Gando:2012zm} independently), and (c)$^{76}$Ge (combined limit from Gerda, IGEX, HDM~\cite{Agostini:2013mzu}). The vertical arrows aim to emphasize the range currently probed with each isotope.  The horizontal, hashed grey band indicates the range of limits on $m_{\beta\beta}$ expected from CUORE assuming its target 90\%C.L lower limit half-life sensitivity of $9.5\times 10^{25}$\,yr is attained.}
%
%
\label{fig:field_comparison}
\end{figure}
To  estimate systematic uncertainties we perform a large number of pseudo-experiments with zero and non-zero signals. We find the bias on $\Gamma_{0\nu}$ from the UEML analysis is negligible.  To estimate the systematic error of the lineshape choice we repeat the analysis of each pseudo-experiment with single- and triple-Gaussian models
and study the deviation of the best-fit decay rate from the posited decay rate as a function of the latter.  
Similarly, we propagate the $5\%$ uncertainty on $\alpha_{\sigma}(Q_{\beta\beta})$, \blue{the $0.09$~keV energy scale uncertainty}, and the choice of zeroth-, \mbox{first-,} or second-order polynomial for the 
background. 
\begin{table}[h]
\caption{Systematic uncertainties on $\Gamma_{0\nu}$ for zero signal (Additive) and as a percentage of nonzero signal (Scaling).}
\label{tab:sysError}
\begin{center}
  \begin{tabular}{lll}
    \hline \hline
\rule{0pt}{2.5ex}    & Additive ($10^{-24}~\rm{yr^{-1}}$) & Scaling (\%)\\
    \hline
   \blue{Lineshape} & {\BBlessShiftFromLineShape} & {\ScalingBBlessShiftFromLineShape}\\
\blue{Energy resolution} & {\BBlessShiftFromEnergyResolution} & {\ScalingBBlessShiftFromEnergyResolution}\\
Fit bias & {\BBlessShiftFromFitBias} & {\BBlessShiftFromFitBiasScaling} \\
\blue{Energy scale}  & {\BBlessShiftFromEnergyScale} & {\ScalingBBlessShiftFromEnergyScale}\\
    \blue{Bkg function}  & {\BBlessShiftFromBkgShape} & {\ScalingBBlessShiftFromBkgShape} \\
    \hline
Selection efficiency &\multicolumn{2}{c}{\BBlessSystErrorFromCutEfficiencies}\\
\hline \hline
\end{tabular}
\end{center}
\end{table}

We combine our data with a {\QinoExposureTeOneThirty} exposure of $^{130}$Te from Cuoricino~\cite{Andreotti:2010vj}. The exposure-weighted mean and RMS
FWHM energy resolution of the detectors were {\QinoMeanEnergyResolution} and {\QinoRMSEnergyResolution}, respectively; the ROI background index was {\QinoROIBkgIndex}.  We report the profile likelihoods in Fig.~\ref{fig:Profile_NLL}. 
The combined Bayesian $90\%$~C.L.\ limit is \mbox{$T_{1/2}^{0\nu}>$~{\CombinedTeOneThirtyBBlessHalflife}}, which is the most stringent limit to date on this quantity.  
For comparison,  the 90\%~C.L. frequentist limits~\cite{bib:Rolke} are $T_{1/2}^{0\nu}>$~{\QZeroTeOneThirtyBBlessHalflifeFreq} for CUORE-0 only, and \mbox{$T_{1/2}^{0\nu}>$~{\CombinedTeOneThirtyBBlessHalflifeFreq}} for the combination with Cuoricino.

We interpret our Bayesian combined limit in the context of models for {\BBless}~decay mediated by light Majorana neutrino exchange using the phase-space factors from~\cite{Kotila:2012zza}, the most recent
nuclear matrix element (NME) calculations for a broad range of models~\cite{Barea:2015kwa,Simkovic:2013qiy,Hyvarinen:2015bda,Menendez:2008jp,Rodriguez:2010mn},
and assuming $g_A \simeq 1.27$ for the axial coupling constant.  The resulting range for the 90\% C.L. upper limit on the effective Majorana mass is \mbox{$m_{\beta\beta}<$~{\CombinedMbbRangeExclSMandPHFB}}; for ease of  comparison with limits from other isotopes in the field (Fig.~\ref{fig:field_comparison}) this range excludes Ref.~\cite{Neacsu:2014bia}. Including the latter NME, the range extends to $m_{\beta\beta}<$~{\CombinedMbbRangeAllNME}.

In summary, CUORE-0 finds no evidence for {\BBless} decay of
$^{130}$Te and, when combined with Cuoricino, achieves the most stringent limit to date on this
process.  Benefiting from lower background, improved
energy resolution, and higher data-taking efficiency, CUORE-0
surpassed the sensitivity of Cuoricino in 
half the runtime.   

The CUORE Collaboration thanks the directors and staff of the Laboratori Nazionali del Gran Sasso and our technical staff for their valuable contribution to building and operating the detector.  \blue{The authors thank J. Feintzeig, L. Gladstone, P. Mosteiro, and V. Singh for carefully reviewing the manuscript, M. Nastasi for preparing the $^{60}$Co calibration sources, and F.~Iachello for helpful discussions concerning the NME literature.}
This work was supported by the Istituto Nazionale di
Fisica Nucleare (INFN); the National Science
Foundation under Grant Nos. NSF-PHY-0605119, NSF-PHY-0500337,
NSF-PHY-0855314, NSF-PHY-0902171, NSF-PHY-0969852, NSF-PHY-1307204, and NSF-PHY-1404205; the Alfred
P. Sloan Foundation; the University of Wisconsin Foundation; and Yale
University. This material is also based upon work supported  
by the US Department of Energy (DOE) Office of Science under Contract Nos. DE-AC02-05CH11231 and
DE-AC52-07NA27344; and by the DOE Office of Science, Office of Nuclear Physics under Contract Nos. DE-FG02-08ER41551 and DE-FG03-00ER41138. 
This research used resources of the National Energy Research Scientific Computing Center (NERSC).

\end{document}